\documentclass{aastex631}

\graphicspath{{./}{figures/}}

\usepackage[normalem]{ulem} 
\usepackage{rotating}
\usepackage{graphicx} 

\usepackage{float}

\usepackage{subfigure}

\def\beginrotate{\begin{Sbox}}
\def\endrotate{\end{Sbox}\rotatebox{-90}{\TheSbox~~}}

\usepackage{textcomp} 
\newcommand{\qdist}[1]{\ifmmode\langle#1\rangle\else\textlangle#1\textrangle\fi}

\usepackage{booktabs}
\newcolumntype{N}{>{\centering\arraybackslash}m{.5in}}
\newcolumntype{G}{>{\centering\arraybackslash}m{2in}}

\begin{document}

\title{First detection of CO$_2$ emission in a Centaur: JWST NIRSpec observations of 39P/Oterma}

\correspondingauthor{Olga Harrington Pinto} \email{oharrington@knights.ucf.edu}

\author[0000-0002-2014-8227]{O. Harrington Pinto}
\affiliation{Department of Physics, University of Central Florida}

\author[0000-0002-6702-7676]{M. S. P. Kelley}
\affiliation{Department of Astronomy, University of Maryland, College Park, MD 20742, USA}

\author[0000-0002-2662-5776]{G. L. Villanueva}
\affiliation{NASA Goddard Space Flight Center}

\author[0000-0003-4659-8653]{M. Womack}
\affiliation{Department of Physics, University of Central Florida}
\affiliation{National Science Foundation}

\author[0000-0003-0194-5615]{S. Faggi}
\affiliation{Department of Physics, American University}
\affiliation{NASA Goddard Space Flight Center}

\author[0000-0002-0622-2400]{A. McKay}
\affiliation{Department of Physics and Astronomy, Appalachian State University, Boone, NC, 28608, USA}

\author{M. A. DiSanti}
\affiliation{Solar System Exploration Division, Planetary Science Laboratory Code 693, NASA}
\affiliation{Goddard Center for Astrobiology, NASA}

\author[0000-0003-1800-8521]{C. Schambeau}
\affiliation{Florida Space Institute, University of Central Florida}
\affiliation{Department of Physics, University of Central Florida}

\author[0000-0003-1156-9721]{Y. Fernandez}
\affiliation{Department of Physics, University of Central Florida}
\affiliation{Florida Space Institute, University of Central Florida}

\author[0000-0001-9542-0953]{J. Bauer}
\affiliation{Department of Astronomy, University of Maryland, College Park, MD 20742, USA}

\author[0000-0002-4230-6759]{L. Feaga}
\affiliation{Department of Astronomy, University of Maryland, College Park, MD 20742, USA}

\author{K. Wierzchos}
\affiliation{Lunar and Planetary Laboratory, University of Arizona}




\begin{abstract}



Centaurs are minor solar system bodies with orbits transitioning between those of Trans-Neptunian Scattered Disk objects and Jupiter Family comets.  39P/Oterma is a frequently active Centaur that has recently held both Centaur and JFC classifications and was observed with the JWST NIRSpec instrument on 2022 July 27 UTC while it was 5.82 au from the Sun.  For the first time, CO$_2$ gas emission was detected in a Centaur, with a production rate of Q$_{CO_2}$ = (5.96 $\pm$ 0.80) $\times$ 10$^{23}$ molecules s$^{-1}$.  This is the lowest detection of CO$_2$ of any Centaur or comet. CO and H$_2$O were not detected down to constraining upper limits.  Derived mixing ratios of Q$_{CO}$/Q$_{CO_2}$ $\leq$2.03 and Q$_{CO_2}$/Q$_{H_2O}$ $\geq$0.60 are consistent with CO$_2$ and/or CO outgassing playing large roles in driving the activity, but not water, and show a significant difference between the coma abundances of 29P/Schwassmann-Wachmann 1, another Centaur at a similar heliocentric distance, which may be explained by thermal processing of 39P's surface during its previous Jupiter-family comet orbit.  To help contextualize the JWST data we also acquired visible CCD imaging data on two dates in July (Gemini North) and September (Lowell Discovery Telescope) 2022.  Image analysis and photometry based on these data are consistent with a point source detection and an estimated effective nucleus radius of 39P in the range of $R_{nuc}= $2.21 to 2.49~km.

\end{abstract}

\keywords{Centaurs, Comets, Asteroids, JWST}


\section{Introduction} \label{sec:intro}



Centaurs are solar system bodies whose orbits are entirely contained in the giant planet region and are considered to be a transitional state for trans-Neptunian objects (TNOs) evolving into Jupiter-family comets (JFCs).  Their orbits are gravitationally perturbed by Jupiter and to a lesser extent, Saturn \citep{Horner2004, Gladman2008, Liu2019, Sarid2019}.  By our definition, their perihelia are larger than Jupiter's with semi-major axes smaller than Neptune's.  Since many Centaurs are active while still so far from the Sun, observations of their dust and gas comae can give insight into evolutionary processes and serve as powerful tools for tracing primitive material from the formation of the solar system \citep{tegler2003, Lisse2022}.

 
At least 42 Centaurs have been identified as active based primarily on their production of a visible dust coma \citep{Bauer2003, Jewitt2009, Epifani2018,wong2019, Chandler2020, steckloff2020, FM2021}.\footnote{See also the regularly updated list of Cometary Centaurs maintained by Y. R. Fernández \url{https://physics.ucf.edu/~yfernandez/cometlist.html}.} Unlike comets that develop comae as they approach the Sun, most active Centaurs generate comae only occasionally, with the exception of the continuously active 29P/Schwassmann-Wachmann 1 \citep{TR2008, Kossacki2013, Sarid2019, Harris2021}.  Active Centaurs have small diameters and have had recent ($\lesssim10^3$~yr) reductions of their perihelion distances leading to lower perihelion distances \citep{Jewitt2009} and entering their current orbital regions (closer to Jupiter), relatively recently \citep{fernandez2018}. Estimates of mass loss rates from observations of active Centaurs range from 4.3 kg s$^{-1}$ to 5100 kg s$^{-1}$ (excluding upper limits) \citep{Jewitt2009, Fraser2022}.

The drivers of this activity remain largely unknown although the sublimation of icy volatiles, such as CO and CO$_2$, and other physical mechanisms may play a role in generating a dust coma \citep{capria2000, Womack2017, wong2019, Lilly2021}.  Another possible source of activity is the crystallization process of amorphous water ice (AWI), which can happen as far out as $r$ $\sim$16 au for an inactive comet-like body \citep{GL2012}.  Although current evidence for AWI in comets is largely based on indirect observational evidence \citep{Prialnik2022}, AWI is expected to be abundant in Centaurs given the conditions that most of them have experienced \citep{Jenniskens1994, Lisse2022}. 




\begin{figure}
\begin{center}
\includegraphics [width=0.8\linewidth] {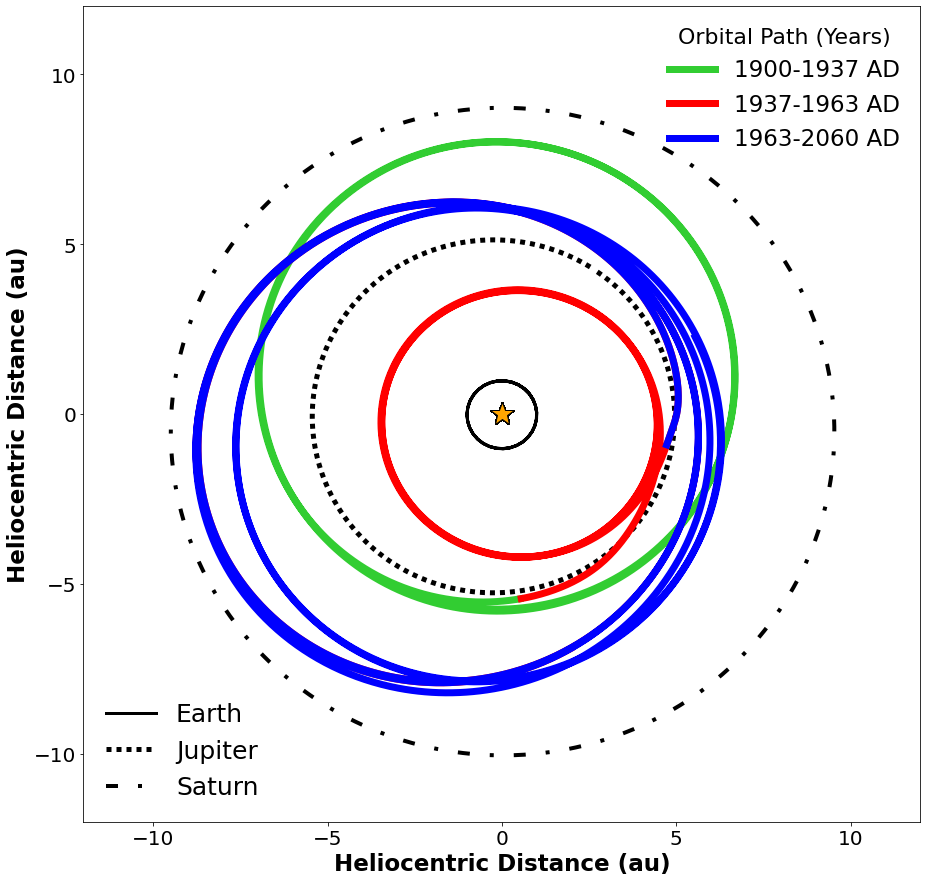}
\caption{ \label{fig:39Porbit} 39P/Oterma's orbit from the years of 1900 to 2060, divided into three periods, is shown here to highlight the drastically varying orbital path it has and will be taking.  The time period from 1900 to 1937 is green, the period from 1937 to 1963 is in red, and the period from 1963 to 2060 is blue.  The orbit of 39P changed significantly from 1943 to 1963, moving from a JFC designation, and a tisserand parameter (T$_J$) that is between 2 and 3, perihelia (q) $<$ 5.2 au, and aphelia (Q) $<$ 7 au \citep{Sarid2019} with $q$ = 3.4 au to a Centaur, with an orbit entirely contained in the giant planet region, with $q$ = 5.5 au, due to a close encounter with Jupiter in 1963.  
The rapid change in 39P's orbit was accompanied by a significant change in solar heating, especially to the outer layers of the nucleus.  This figure was generated using the {\it REBOUND} Python package \citep{rein-2012}.  The progression of time is in the counter clockwise direction.
}
\end{center}
\end{figure} 

39P/Oterma (39P) is an active Centaur that in the last 90 years has transferred from a Centaur orbit to a closer-in Jupiter-family comet orbit and then outward again to Centaur status (Figure \ref{fig:39Porbit}).  It was discovered in 1943 by Liisi Oterma with a visual magnitude of $m \sim$ 15 \citep{marsden1962,Kronk2009}, when it had just passed through perihelion at $r \sim$ 3.4 au and was at $\Delta$ = 2.5 au.  It has often been reported with a non-stellar physical appearance, and in May 1943 it showed up on reflector plates as a star-like object surrounded by a small ring of nebulosity similar to a planetary nebula with a strong central condensation \citep{1943Herbig}.  

Shortly before its discovery, 39P had a close approach to Jupiter in 1937, with a minimum orbit intersection distance, MOID, of 0.16 au, which shortened the orbital period from 18 years to 8 years, transitioning it from a Centaur to a JFC orbit.  Orbital simulations of the Centaur's astrometry show that it had several approaches to Jupiter and Saturn in recent centuries that eventually shortened the perihelion to 3.4 au which it maintained for about 30 years \citep{koon2000}.  Figure \ref{fig:39Porbit} illustrates how the orbit evolved until 1963, when 39P had another close approach to Jupiter, with MOID = 0.1 au that increased its perihelion to 5.5 au, and thereby returned 39P to Centaur status \cite[e.g.][]{Jewitt2009,Kronk2009}.  During 1942 to 1962, 39P was dimmer than $m \sim$ 18 and occasionally exhibited a short tail \citep{marsden1962}.  However, in its 1950 passage, 39P was seen to be its brightest with a $m \sim$ 14.5 \citep{vanBiesbroeck1951,Kronk2009}.  No mention of filters was made in the cited text, so we assume this is a visual magnitude.  39P was not seen from 1962 until 1998, when it was recovered and showed weak activity in 2002 with $r \sim$22~mag \citep{fernandez2001, Jewitt2009}.  39P was not seen again from 2002 until 2019, when it was recovered, showing what could be a compact coma and/or an elongated nucleus with $r \sim$24~mag \citep{schambeau2019}. 



Water, CO, and CO$_2$ are the most abundant volatile species in comets and measurements of these species' abundance in active Centaurs are needed to constrain models of solar system formation and evolution. Unfortunately, due to their distance and relatively weak activity, very few such observations are recorded for Centaurs (see \citet{Fraser2022}).  CO$^+$ was the first gaseous species observed in a Centaur and its repeated detection in 29P/Schwassmann Wachmann 1 (29P) indicated that CO was being produced in large amounts \citep[e.g.,][]{cochran1980, Larson1980, Cochran1991}.  Although CO$^+$ can be produced by photoionization of CO in comets, this process is inefficient at Centaur distances and CO$^+$ column densities and variability can be better explained by solar-wind proton impact onto CO which strongly depends on solar-wind particle velocities \citep{Cochran1991, Jockers1992, Ivanova2019, Wierzchos2020}. 
CO has been detected regularly in 29P's coma \citep[e.g.,][]{senay1994,festou2001,Paganini2013, Womack2017, Wierzchos2020, bockelee2022}, and single detections were reported for the less-active Chiron \citep{womack1999} and Echeclus \citep{weirzchos2017}.  CO was searched for, but not detected, in the coma of P/2019 LD2 (ATLAS), an object transitioning from a Centaur to a JFC orbit \citep{Kareta2021}.  H$_2$O emission was detected in 29P and was attributed to the sublimation of icy grains \citep{oot2012,bockelee2022}. 

CO$_2$ emission is much more difficult to measure than CO for several reasons and has only been directly measured in 26 comets, none of which is a Centaur \citep{HP2022}.  Most Centaurs have spent a large portion of their orbit in a region that is too warm to keep surface layers of CO and CO$_2$ ice \citep{Jewitt2009} and activity that we see in Centaurs that are as far out as $\sim$ 15 au could be the result of a subsurface pocket of CO$_2$ \citep{Lilly2021}.  NASA's newly commissioned James Webb Space Telescope (JWST), a 6.5-m infrared optimized space telescope \citep{gardner2023}, is needed to advance our understanding of CO$_2$ in Centaurs since it is the only current telescope that has the capability to observe this molecule independently from other molecules.

Here we present spectra of 39P obtained with the JWST Near-infrared Spectrograph (NIRSpec) instrument over the wavelength range of 0.6--5.3 microns when 39P was 5.82 au from the Sun.  Supporting ground-based optical observations of 39P are also presented from the Gemini North Observatory and Lowell Discovery Telescope using broad-band filters to provide estimates of the dust production and the nucleus's radius near the time of JWST observations. 

\section{Observations and Results} \label{sec:obs}

\subsection{JWST}
\label{subsec:JWST}

JWST was used to observe 39P with the NIRSpec \citep{Jakobsen2022,boker2023} 
instrument in Integral Field Unit (IFU) PRISM mode over a 3$^"$ × 3$^"$ square region, each pixel on the detector covering 0.01 arcsec$^2$.  Spectra were obtained over the wavelength range 0.6--5.3 microns encompassing emissions from the strong fundamental vibrational bands of H$_2$O, CO$_2$, and CO.  The spectral resolution for NIRSpec IFU PRISM is R $\sim$ 100 \citep{boker2023}.  The data were processed with the JWST pipeline version v1.10.1 \citep{bushousehoward20237826579} and JWST Calibration Reference Data System (CRDS) context file number 1089 and can be found in MAST: \dataset[10.17909/72k4-ec72]{http://dx.doi.org/10.17909/72k4-ec72}.

\begin{deluxetable*}{ccccccccccc}[h!] \tablecaption{Observing Details \label{tab:photom-obs}}
\tablecolumns{11}
\tablewidth{0pt}
\tablehead{
\colhead{Telescope} &
\colhead{UT Date} &
\colhead{Time$^a$} &
\colhead{$r$} &
\colhead{$\Delta$} &
\colhead{$\alpha$} &
\colhead{Seeing} &
\colhead{Filter} &
\colhead{Total Exp. Time} &
\colhead{Airmass$^b$} &
\colhead{$mag.$}\\      
\colhead{} &
\colhead{(YYYY-MM-DD)} & 
\colhead{(UTC)} &
\colhead{(au)} &
\colhead{(au)} &
\colhead{$(^{\circ})$} &
\colhead{(arcsec)} & 
\colhead{} & 
\colhead{(seconds)} &
\colhead{} &
\colhead{} 
}

\startdata
JWST & 2022-07-28 & 21:49:26 & 5.815 & 5.485 & 9.872 & $\sim$ 0\farcs1 & CLEAR & 1021.222 & N/A & N/A \\
Gemini & 2022-07-28 & 13:35:24 & 5.815 & 5.490 & 9.769 & 0\farcs65 & $r'$ & 480 & 1.107 & 23.45 $\pm$ 0.03 \\
Lowell Discovery & 2022-09-18 & 05:16:39 & 5.786 & 4.852 & 4.275 & 1\farcs1 & $r'$ & 300 & 1.67 & 22.65 $\pm$ 0.05 \\
\enddata

\tablecomments{\footnotesize{$^a$ UTC at start of image sequence.\\
$^b$ Mean airmass of 39P during exposure.}}
\end{deluxetable*}

Wavelength calibration used an Argon emission lamp from the first cryogenic performance testing, before JWST was deployed \citep{Birkmann2011}.\footnote{More in depth details behind the wavelength calibration can be found in \citet{Birkmann2011}, \citet{de2012}, \citet{Boker2022}, and \citet{Jakobsen2022}.} Flux calibration was accomplished in Stage 2 of the JWST calibration pipeline, where the combination of the scalar conversion factor and the 2-D response values was applied to the science data \footnote{More details of the JWST calibration pipeline can be found at https://readthedocs.org/projects/jwst-pipeline/downloads/pdf/latest/}.  A few important things to note are that the absolute calibration is currently limited to 10\%, and that the data from JWST are presently calibrated to an aperture radius of 0.4\arcsec.  We have accounted for these corrections in Table \ref{tab:chemcomp}.  The aperture correction is based on observations of the G2V star SNAP-2 from the NIRSpec commissioning program 1128.  The correction varies with wavelength, and we calculate multiplicative factors of, 1.08, 1.13, and 1.16 at 2.66, 4.26, and 4.67~\micron, respectively.

\begin{figure}
\begin{center}
\includegraphics [width=0.5\linewidth] {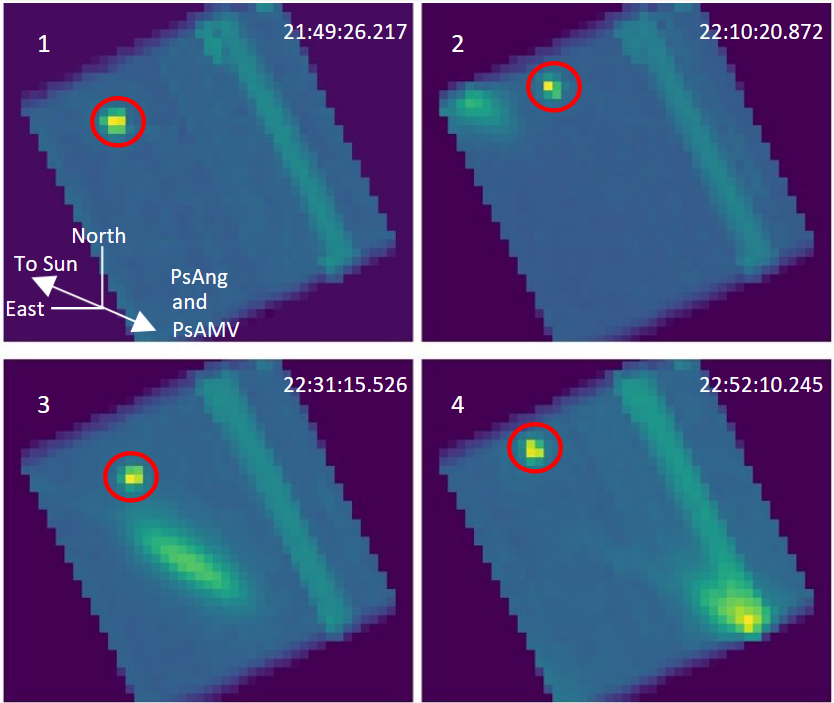}
\caption{ \label{fig:39PMedians} These images of 39P are extracted from each of the four NIRSpec datacubes at the median over the wavelength axis of of wavelengths from the datacubes, $\sim$2.95 $\mu$m. The data were all taken on 2022 July 27 and the UT starting exposure time is provided in the upper right of each image panel.  A 0.5 arcsecond diameter aperture (denoted with a red circle and corresponding to a projected diameter of 1994.5 km at the Centaur's distance from JWST) was used to extract the 39P spectrum.  A background galaxy was imaged in frames 2--4, but does not contaminate 39P's coma in the aperture used.  The images are oriented with north up and east to the left.  The projected Sun-comet vector (PsAng) and projected negative orbital velocity vector (PsAMV) are shown, obtained using JPL's Horizons.  These vectors are used as indicators of the potential tail direction.
}
\end{center}
\end{figure}

On 2022 July 27 UTC, JWST observed 39P, when the Centaur was 5.82 au from the Sun, 5.50 au from JWST, and at a phase (Sun-target-telescope) angle of 9.87\degr.  Four dither positions where obtained with an effective exposure time of 1021.222 seconds (based on EFFEXPTM from the header of the datacubes).  The sequence start, mid, and stop times were 21:49:26, 22:29:26, and 23:09:26 UTC, respectively.  The telescope tracked the Centaur at the predicted ephemeris obtained from JPL Horizons.  The dither offsets were $\sim$0\farcs4 from the 4-point dither setting.  This is done to improve the spatial and spectral reconstruction without gaps or detector artifacts.  Four dithers (spectral datacubes) of 39P, each with a total exposure duration of 1035.811 seconds (about 17 minutes, based on DURATION from the header of the datacubes) were obtained in the span of about 90 minutes.  The difference between the total exposure time is the integration time multiplied by the number of integrations while the effective exposure time is corrected for dead time and lost time. Dedicated background frames were obtained 24 minutes after the 39P observations, the distance between the apparent position of the sky between the first observation of 39P and the last observation of the offset of 39P is 0.016 degrees.  In the first observation 39P stands alone, except for a stripe feature toward the right which is the result of one of the micro-shutters suffering an electrical short and so it begins to glow.  This stripe did not overlap with the aperture used for any of the 39P data analysis.  In the other three frames a background galaxy appears to pass through the field but it also does not overlap with the Centaur's position on the sky (Figure \ref{fig:39PMedians}).

\begin{figure}
\begin{center}
    \includegraphics[width=0.85\textwidth]{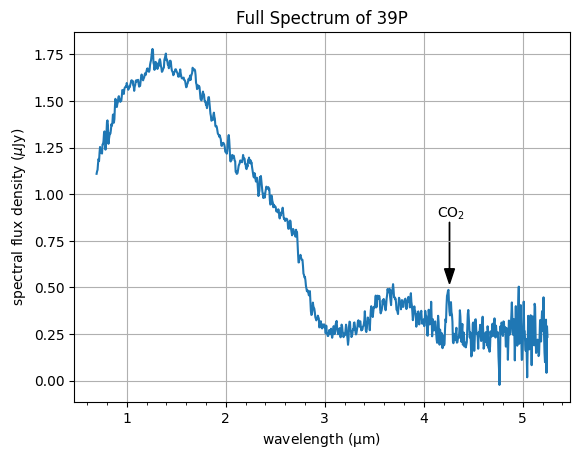}\hfill
    \caption{JWST NIRSpec spectrum of 39P is shown, where the bump at 4.26 $\mu$m is the detection of CO$_2$.  The CO and H$_2$O signatures are not detected.  CO and H$_2$O would be seen at 4.67 $\mu$m and 2.66 $\mu$m, if their emissions were stronger.}\label{fig:39fullspec}
\end{center}
\end{figure}

After running the data through the JWST Calibration pipeline, which defaults the orientation to North being up and East being left (see Figure \ref{fig:39PMedians}), then we rotated to align the detector orientation with the NIRSpec IFU plane and shifted the last three dithers to align in RA and Dec coordinates with the first dither and to account for the NIRSpec reference axis.  We used a photometric mask of 39P and a small area around it to remove background signal.  All non-zero pixels were used to remove the median flux density of the frame and any outlying pixels (like that of the passing galaxy).  The annulus used was 0.3 arcsec (difference between the radius of the background mask and the radius of the 39P mask) and no gas was seen outside of the 39P mask. We then used the median of the four dithers to get our final spectral cube for 39P, which also removed any hot pixels.  We use an aperture diameter of 5 pixels across as seen in Figure \ref{fig:39PMedians}, which corresponds to a projected diameter of 1994.5 km for 39P at the time of these observations.  To extract the 1-D spectrum from the 2-D IFU datacube this technique was applied to each slice of the datacube, to then construct a 1-D spectrum from the slices of the final datacube at different wavelengths.  The NIRSpec calibration pipeline corrects for the movement of 39P down to an accuracy of 50 mas, so in our analysis we assumed that the centroid did not change. In Figure \ref{fig:39fullspec}, the spectrum is dominated by reflected light, a broad 3-$\mu$m absorption band, and a 4.26 $\mu$m emission band.  Contributions from the dust continuum were removed from the spectrum by subtracting a second or third order polynomial function fit to regions surrounding each molecular signature.  See Section \ref{subsec:ice} for detailed analysis about possible ice or dust features in the spectrum.

\begin{figure}
\begin{center}
\includegraphics [width=0.5\textwidth] {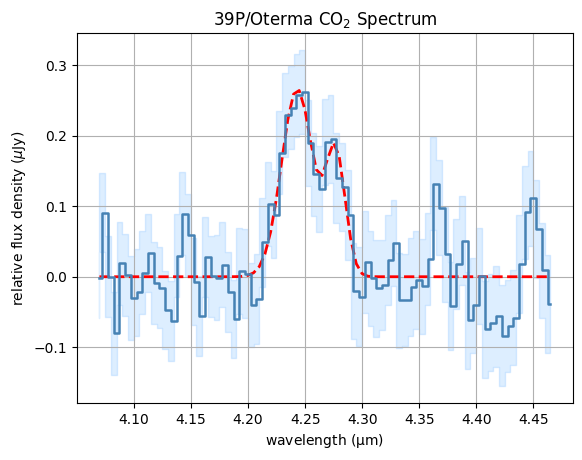}
\caption{ \label{fig:39PSpecfit} JWST NIRSpec spectrum of CO$_2$ emission in 39P after the dust continuum has been subtracted.  This is the first detection of CO$_2$ emission in a Centaur.  A double Gaussian fit is superimposed on the double peaks of the CO$_2$ ($\nu_3$ band) which we used to measure the integrated flux density.  We derive a production rate of Q$_{CO_2}$ = 5.96 $\pm$ 0.80 $\times$ 10$^{23}$ molecules sec$^{-1}$.  The uncertainties of the relative flux densities associated with each spectral channel is shown by the region highlighted in light blue.
}
\end{center}
\end{figure}

The integrated flux density for CO$_2$ was determined by removing the continuum and fitting a double Gaussian over the bandwidth range 4.07 to 4.47 $\mu$m.  The uncertainty was calculated using the standard deviation of the fit.  We measured a 7-sigma detection of CO$_2$'s integrated flux in 39P based on emission in the 4.3 micron $\nu_3$ vibrational band (see Figure \ref{fig:39PSpecfit}).

\begin{figure}
  \begin{minipage}{\textwidth}
    \includegraphics[width=0.5\textwidth]{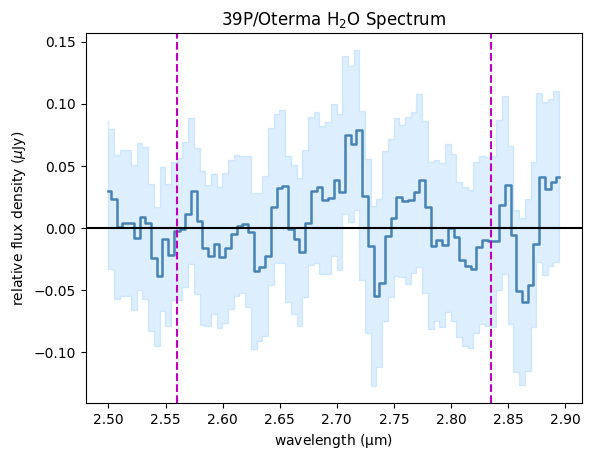}\hfill
    \includegraphics[width=0.5\textwidth]{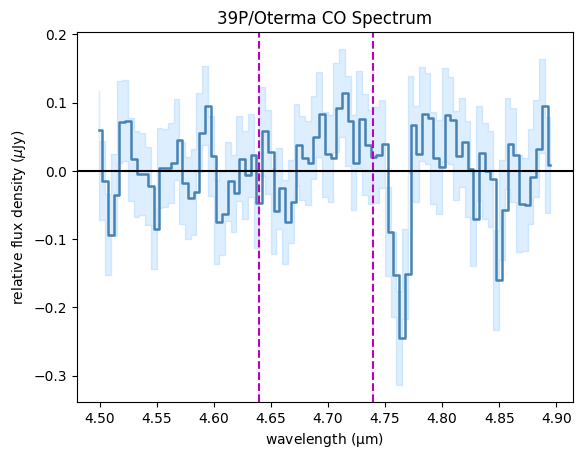}\hfill
    \caption{H$_2$O (left) and CO (right) spectra of 39P with the continuum subtracted.  Neither volatile was clearly detected.  Here we get the upper limits of 10.0 $\times$ 10$^{23}$ molecules s$^{-1}$ and 12.1 $\times$ 10$^{23}$ molecules s$^{-1}$, respectively.  In both spectra, the wavelengths from Table \ref{tab:chemcomp} where we would expect to see the molecular peak are located between the dashed purple lines, these purple lines mark the beginning and end of the region that was integrated to get the upper limits of the production rates. The noise surrounding the purple dashed lines is comparable to the molecular signatures themselves.  The uncertainties of the relative flux densities associated with each spectral channel is shown by the region highlighted in light blue.\label{fig:39PH2OSpec}}
  \end{minipage}
\end{figure}

No strong signal is detected for either H$_2$O or CO, so upper limits were determined.The continuum for H$_2$O was obtained from $\sim$2.5–2.9 $\mu$m where a polynomial fit well against the spectra of 39P.  However, there are not enough detections of water ice in comets and Centaurs at these wavelengths to form a basis for understanding how well a polynomial will work in the general situation. Upper limits were calculated integrating the flux from 2.56–2.83 $\mu$m for H$_2$O and 4.64–4.72 $\mu$m for CO,  following the method described in \citet{oot2010}.  Then, the integrated flux value is used to calculate the upper limit to the production rates, using Equations \ref{eq:cdensity} and \ref{eq:prodrate}.  These production rates are treated as upper limits since the regions around the wavelengths of these volatiles are very noisy. The regions are seen between the dashed lines in the spectra in Figure \ref{fig:39PH2OSpec}.  Measured relative flux densities, and upper limits, are listed in Table \ref{tab:chemcomp}.  To compare 39P against the comets from \cite{HP2022} the fluorescence efficiencies from \citet{1983crovenc} were used when calculating the production rates.  \cite{HP2022} compares proxy values of CO$_2$ and CO that were obtained using various methods (like [O I], CO Cameron bands, and IR measurements from Spitzer).\\

\subsection{Gemini North}
\label{subsec:Gemini}

Thirteen hours after the JWST NIRSpec data were obtained, observations were acquired from the 8.1-m Gemini North telescope using the Gemini Multi-Object Spectrograph (GMOS).  Four images were acquired in the GMOS $r'$ (620 nm) filter using an exposure time of 120 seconds each.  The detector pixels were binned 2$\times$2 resulting in a $\sim$ 0.16$''$ pixel scale.  Table \ref{tab:photom-obs} summarizes the observing circumstances. 

All GMOS images were reduced, including bias subtraction, flat-field correction, and mosaicking of the three CCD images using the Gemini {\it DRAGONS} software \citep{dragons-2019}.  Individual images were then processed using our Python-based GMOS reduction and calibration pipeline which includes (1) cosmic-ray removal performed utilizing the LACosmic technique \citep{van-dokkum-2001} as implemented in ccdproc \citep{craig-2017}, (2) stacking of individual images in 39P's non-sidereal frame and the sidereal frame, and (3) a Pan-STARRS-based flux calibration using field stars from the GMOS central CCD (which included 39P) stacked in the sidereal frame.  The Pan-STARRS magnitudes used for calibrations are from the PS1 DR2 MeanObject tables reported in the AB magnitude system \citep{magnier-2013}. 

\begin{figure}
\begin{center}
    \includegraphics[width=0.45\textwidth]{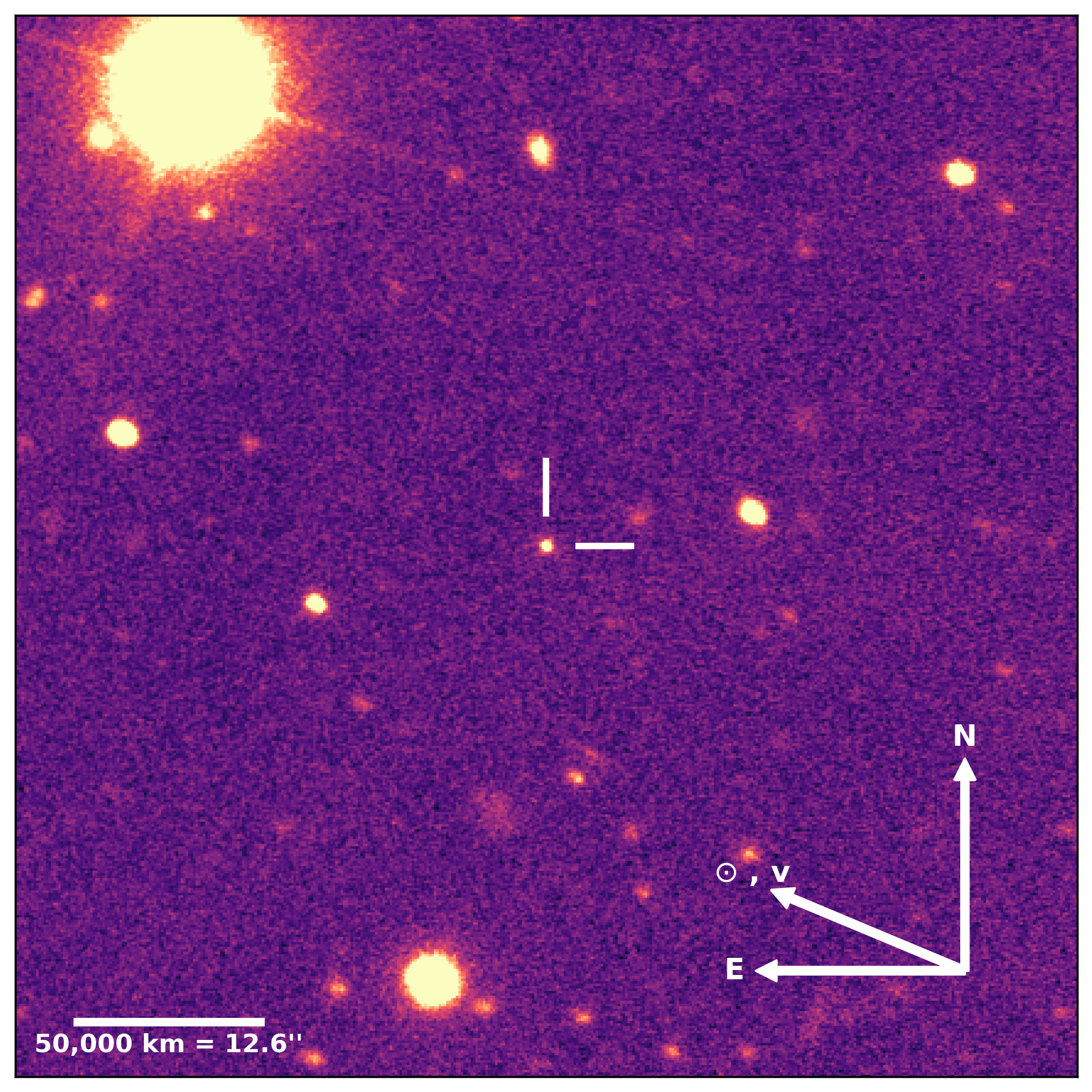}\hfill
    \includegraphics[width=0.45\textwidth]{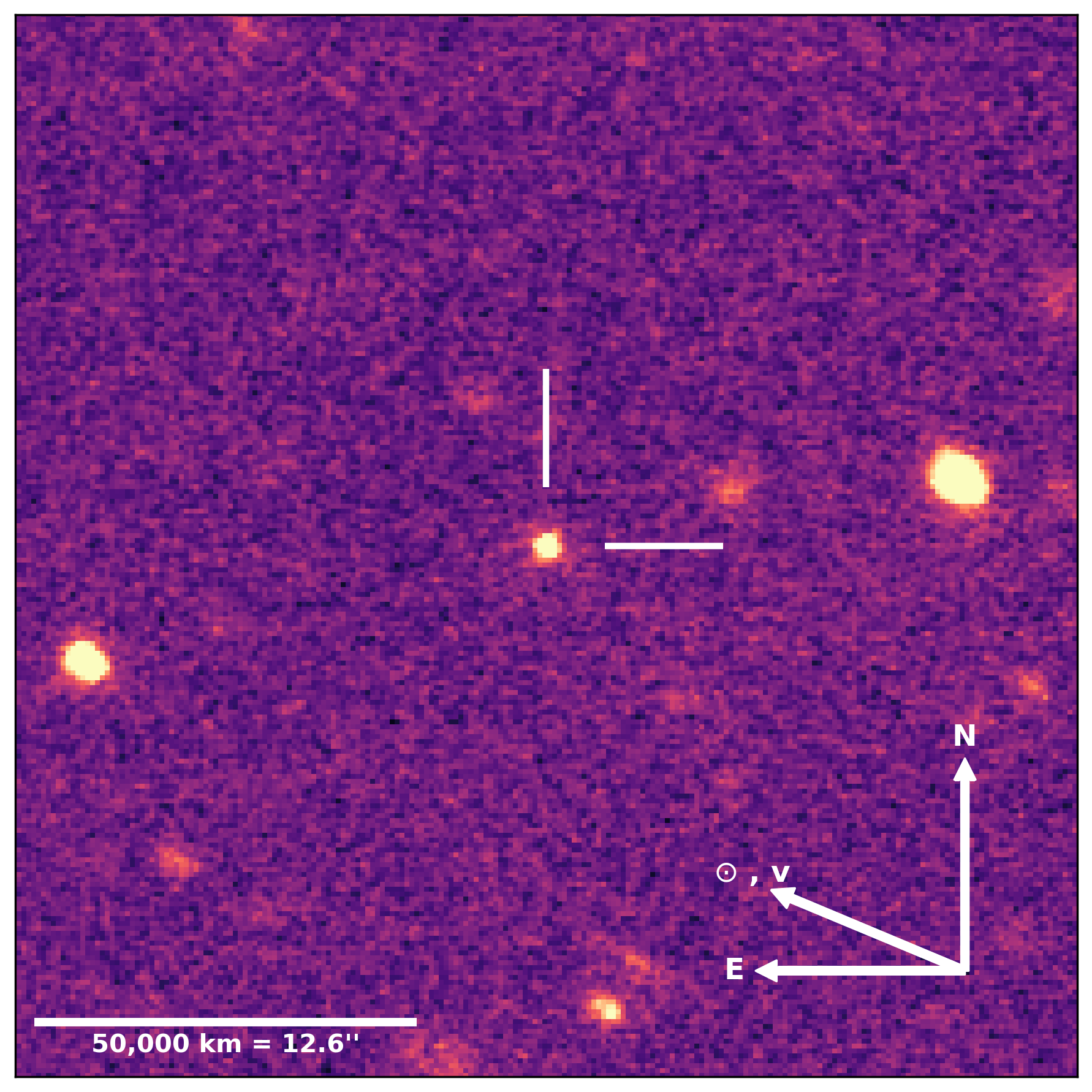}
    \caption{ \label{fig:gemini} (left) Gemini North Telescope image of Centaur 39P obtained using an $r^\prime$ filter on 2022 July 28 UTC.  39P is at the center of the image, marked with two white lines.  The image scale and orientation with the J2000 equatorial reference frame are shown, as well as the projected comet-Sun vector ($\odot$ = 68\degr) and the projected velocity vector ($v$=68\degr).  (right) 2$\times$ zoomed image of 39P taken with Gemini North.}
\end{center}
\end{figure}

Figure \ref{fig:gemini} displays the final non-sidereal stacked GMOS image centered and cropped around 39P.  Corresponding photometry measurements are included in Table \ref{tab:photom-obs}.  Comparisons of radial surface brightness profiles of 39P compared to field stars indicate no profile extension due to a dust coma being detected in the Gemini imaging data.\\





\subsection{Lowell Discovery Telescope}
\label{subsec:LDT}

Images of 39P were taken with the 4.3-m Lowell Discovery Telescope, located near Happy Jack, AZ.  We used the Large Monolithic Imager (LMI) which has a 6144~pix$\times$6160~pix e2v CCD with an unbinned pixel scale of 0\farcs12 for a total field of view of 12\farcm3$\times$12\farcm3.  39P was observed with a Sloan $r^\prime{}$ filter \citep{Fukugita1996}, and the data were bias subtracted and flat fielded with standard techniques.  Three images with 2$\times$2 pixel binning and 300-s exposure times were taken, tracking at the ephemeris rate of the Centaur, with a mid-observation time of 2022 September 18 05:24 UTC.  Table~\ref{tab:photom-obs} summarizes the observing circumstances. 
These observations were taken 57 days after the JWST observations of 39P, but given the small change in heliocentric distance and insolation, $r_{(\mathrm{LDT})}^2/r_{(\mathrm{JWST})}^2$=1.04, they are useful for context.

\begin{figure}
\begin{center}
    \includegraphics[width=0.5\textwidth]{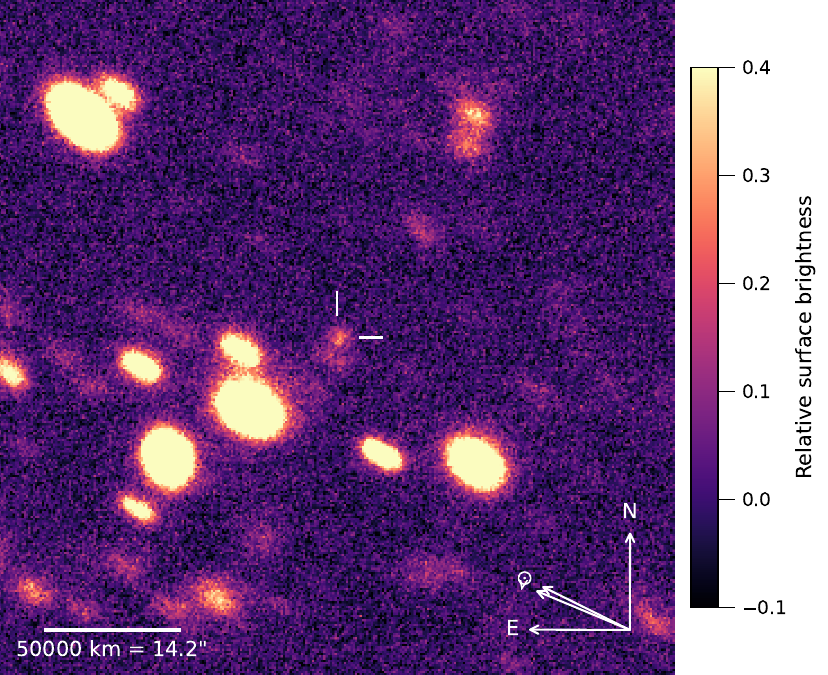}\hfill
    \includegraphics[width=0.5\textwidth]{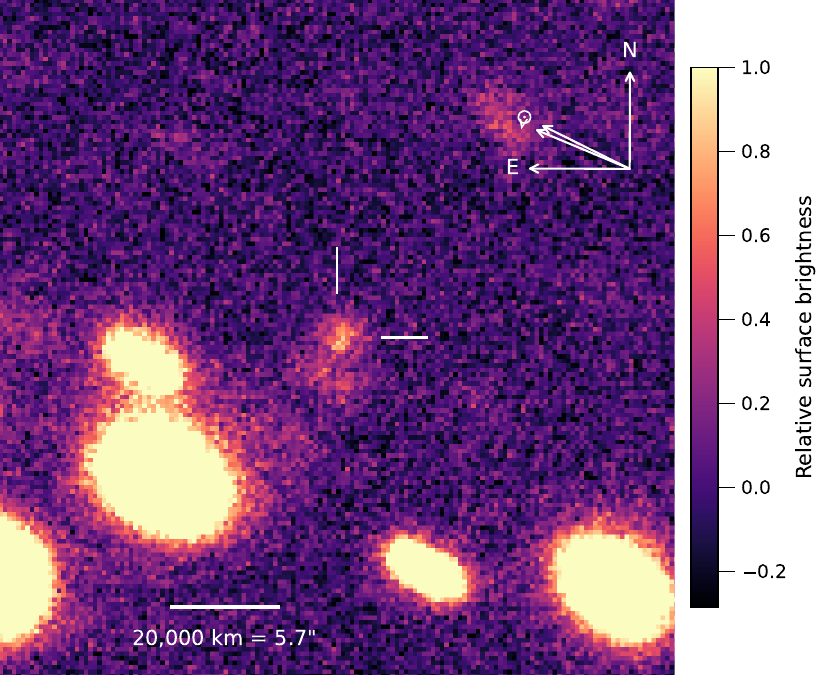}
    \caption{ \label{fig:ldt} (left) Image of Centaur 39P obtained with the 4.3-m Lowell Discovery Telescope using an $r^\prime$ filter on 2022 September 18 UTC.  39P is at the center of the image, marked with two lines.  This image was obtained 57 days after the JWST data but provides useful context.  The image scale and orientation with respect to the J2000 equatorial reference frame are shown, as well as the projected Centaur-Sun vector ($\odot$ = 64\degr) and the projected velocity vector ($v$=68\degr). (right) 2$\times$ zoomed image of 39P taken with LDT.}
\end{center}
\end{figure}

The photometric calibration is based on the Pan-STARRS1 photometric system using the calviacat software with background stars observed throughout the half-night of observing and the ATLAS-Refcat2 photometry catalog hosted at the Mikulski Archive for Space Telescopes \citep{Tonry2012, Tonry2018, Kelley2019-calviacat}.  Finally, we defined the world coordinate system for each image using Astrometry.net and the Gaia DR2 catalog \citep{Lang2010, Gaia2018}.  The mean-combined image is presented in Figure~\ref{fig:ldt}.

Photometry was measured with a 1\farcs2 radius aperture centered on 39P.  The aperture is small compared to the 1\farcs1 seeing for these data, but was necessary to avoid a nearby background object that passed as close as 2\farcs2 from 39P.  The relative rate between 39P and background sources was 1\farcs2 per exposure, so rather than use the background stars for the aperture correction, we used 39P itself (for details, see Appendix~\ref{app:aperture}).  The total flux density from 39P is measured on each of the three images, averaged together, then corrected by a factor of 1.670 to match an effective aperture radius of 2\farcs16, encompassing the full extent of the source.

\section{Discussion} \label{sec:discussion}

\subsection{Af$\rho$ and nucleus size from optical photometry}\label{sec:nuc-size}

The nucleus of 39P is not resolvable through the LDT or Gemini observations, however the flux received by the telescope provided a point-source ``image'' of 39P. Even though 39P was not resolved, we can assume the flux received was from reflected flux from the nucleus’ surface.  We applied standard techniques to estimate an effective spherical nucleus size following \cite{knight-2023}.  The LDT and Gemini observations were acquired to better understand the dust coma environment present during the JWST observations.  Interestingly, no extension beyond a point-source, indicating the absence of a conspicuous dust coma, was present for 39P during either epoch of imaging.  This is most important in the Gemini data, which were acquired approximately one day after the JWST CO$_2$ detection.

The JWST data were also not able to resolve the nucleus of 39P, though based on the size estimates from Gemini an LDT and the aperture used for the JWST spectral analysis we see that there is a gas/ice coma.  From the spatial profiles in Figure \ref{fig:39P_CO2Profile} it does not seem like a diffuse dust coma was present.

Optical depth effects can occasionally affect calculations of the nucleus radius from imaging data, so the grain filling factor was estimated to see how the coma of 39P might be affected by optical depth effects.  The filling factor, $f$ is one way to examine the optical depth effects from the dust since it is the ratio between the optically thick area and the total area \citep{AM2017}.

\begin{equation}
     Af\rho = {{4 \Delta^2 r^2}\over{\rho}} 10^{0.4({m_{Sun}-m_{39P}})}
\label{eq:afp}
\end{equation}

\begin{equation}
     f = {{4 \Delta^2 r^2}\over{A\rho^2}}10^{0.4({m_{Sun}-m_{39P}})}
\label{eq:fillingfactor}
\end{equation}



To get the filling factor, the Af$\rho$ was first calculated using Equation \ref{eq:afp} and the following values: apparent magnitude for 39P from LDT (22.65), the SDSS r' band apparent magnitude of the Sun (-27.08) from \citep{Willmer2018}, the corresponding heliocentric (5.79 au) and geocentric distances (4.86 au), and the aperture radius (7.61 $\times$ 10$^3$ km) for 39P when it was observed with the LDT.  Using the details of the observation with the LDT, the Af$\rho$ was calculated to be 12 cm.  Then, using Equation \ref{eq:fillingfactor}, with the albedo, A, and aperture radius from the LDT observations, the filling factor was determined.  For these calculations we assumed a nominal $r$-band cometary geometric albedo of 0.05.  
39P has a  $f  =$  3.13 $\times$ 10$^{-7}$, 
which is an extraordinarily low value suggesting that the optically thin approximation is appropriate.  Since $f$ is low this means there is a very small cross section of grains that fills the field of view.

The filling factor is an important quantity that allows us to understand the dust comae structure without the necessity of details.  An increased optical depth would reduce the amount of solar radiation reaching the nucleus, and cause the light to be scattered by particles in the coma rather than the nucleus \citep{Afghan2023} and this could make the size estimate of the nucleus a higher value than it really is.

We used traditional methods of bare nucleus analysis \citep{lamy_2004, knight-2023} to estimate an effective spherical radius for 39P's nucleus based on the Gemini and LDT images.  39P had an apparent magnitude of $r$=22.65$\pm$0.05~mag in the LDT optical images and $r$=23.45$\pm$0.03~mag in the Gemini images, quoted for the PS1 r-band, in the AB magnitude system.  Assuming a log-linear phase function with slope 0.046~mag~deg$^{-1}$, we calculate absolute magnitudes of $H_r(1,1,0)$=15.21~mag and $H_r(1,1,0)$=15.48~mag, respectively.  These absolute magnitudes are consistent with the estimate from \cite{Jewitt2009}.  $H_r(1,1,0)$ is a proxy for the size \citep{Peixinho2012} and since Centaurs with $H_r(1,1,0)$ $\geq$ 6.8 could have bimodal color distribution \citep{fraser2012,wong2017} the sizes we determined are indicative of 39P possibly having one of two surface colors, its surface color is either red or grey/neutral colored.  Centaurs have previously been observed to have two distinct color populations \citep{tegler2008,Peixinho2012,wong2017,Liu2019}.  Using these absolute magnitudes we estimate the nucleus radius to be $R_{nuc}= $2.49 $\pm$0.05~km based on the LDT data, and $R_{nuc}= $2.21 $\pm$0.03~km based on the Gemini data.  The measured magnitudes are consistent with that of a bare-nucleus based on a size estimate measured during 39P's last recovery \citep{schambeau2019} and previous observations of 39P \citep{Jewitt2009}.  The radius derived by the LDT data is slightly larger than the one determined with the Gemini observations.  Since possible contamination from a close-by field star in the LDT data was not included in the photometry or in the calculations, this difference in radii could be attributed to 39P possessing an elongated shape that was observed at different angles.\\ 



\subsection{CO$_2$, CO, and H$_2$O column densities and production rates}

We calculated the column densities using 



\begin{equation}
\label{eq:cdensity}
\qdist{N_x} = F_{x}4\pi\Delta^2{\lambda\over{hc}}{r^2\over{g}}{1\over{\pi\rho^2}}
\end{equation}

where \qdist{N$_x$} is the column density of any molecule (represented by x) along the line of sight, F$_{x}$ is the integrated flux density of a molecule measured with NIRSpec \footnote{To convert from the integrated flux density Jy $\mu$m to erg cm$^{-2}$ s$^{-1}$ we multiplied by 3 $\times$ 10$^{-1}$ $\times$ $\lambda_{x}^{-2}$} 
(see Table \ref{tab:chemcomp}), $\Delta$ is the distance between 39P and JWST, $\lambda$ is the wavelength of the observed emission, $h$ is the Planck constant, $c$ is the speed of light, $g$ is the fluorescence efficiency at 1 au (seen as the g-factor in Table \ref{tab:chemcomp}), $r$ is the heliocentric distance in au as a scaling factor for the fluorescence efficiency (no units), and $\rho$ is aperture radius, based on the projected linear scale.


The molecular production rate in molecules s$^{-1}$, Q$_x$, is calculated using 
\begin{equation}
\label{eq:prodrate}
Q_x = \qdist{N_x}2v\rho
\end{equation}

where $v$ is the gas expansion velocity \footnote{We use a gas expansion velocity of 0.24 km s$^{-1}$, based on the gas expansion velocity equation by \citet{delsemme1982} of $v$ = 0.58 $\times$ $r^{-0.5}$ km s$^{-1}$.} and $\rho$ is projected aperture radius at the distance of 39P.  Values for $\qdist{N_x}$ and $Q_x$ are tabulated in Table \ref{tab:chemcomp}.

\begin{deluxetable*}{cccccc}
\renewcommand{\arraystretch}{1.1}
\tabletypesize{\scriptsize}
\tablecaption{Integrated band flux densities, column densities and production rates for 39P/Oterma from 2022 July 27 \label{tab:chemcomp}}
\tablecolumns{6}
\tablewidth{0pt}
\tablehead{
\colhead{Molecule} &
\colhead{$\lambda_x$} & 
\colhead{Integrated Flux} &
\colhead{g} &
\colhead{$\qdist{N_x}$} & 
\colhead{Q$_x$$^a$} \\
\colhead{} &
\colhead{$\mu$m} &
\colhead{10$^{-19}$ erg cm$^{-2}$ s$^{-1}$} &
\colhead{10$^{-4}$ s$^{-1}$} &
\colhead{10$^{10}$ mol cm$^{-2}$} & 
\colhead{10$^{23}$ mol s$^{-1}$}
}
\startdata
    H$_2$O ($\nu_3$) & 2.66 & $\leq$4.63 & 2.85 & $\leq$20.0 & $\leq$10.0 \\
    CO$_2$ ($\nu_3$) & 4.26 & 18.0$\pm$ 1.61 & 28.6 & 12.4$\pm$ 1.12 & 5.96 $\pm$ 0.80\\
    CO v(1–0) & 4.67 & $\leq$2.74 & 2.46 & $\leq$24.2 & $\leq$12.1 \\
\enddata
\small
\tablecomments{\footnotesize{$^a$Uncertainties and upper-limits include a 10\% absolute calibration uncertainty.}}

\end{deluxetable*}

\begin{deluxetable*}{ccccccc}
\renewcommand{\arraystretch}{1.1}
\tabletypesize{\scriptsize}
\tablecaption{Volatiles in Two Active Centaurs: 39P/Oterma and 29P/Schwassmann-Wachmann  1\label{tab:versus}}
\tablecolumns{7}
\tablewidth{0pt}
\tablehead{
\colhead{Centaur} &
\colhead{r} & 
\colhead{Q$_{CO}$/Q$_{CO_2}$} &
\colhead{Q$_{CO_2}$/Q$_{H_2O}$} &
\colhead{Q$_{CO_2}$} &
\colhead{Q$_{CO}$} &
\colhead{Q$_{H_2O}$} \\
\colhead{} &
\colhead{au} &
\colhead{} &
\colhead{} &
\colhead{10$^{23}$ mol s$^{-1}$} &
\colhead{10$^{23}$ mol s$^{-1}$} &
\colhead{10$^{23}$ mol s$^{-1}$}
}
\startdata
    39P & 5.82 & $\leq$2.03 &$\geq$0.60& 5.96 $\pm$ 0.80 & $\leq$12.1 & $\leq$10.0\\
    29P$^a$ & 6.18 & $\geq$83.2 & $\leq$0.05& $\leq$3500 & 291500 & 67800\\
\enddata
\small
\tablecomments{
\footnotesize{$^a$These values are from \citet{HP2022} which used the average of the reported values obtained by \citet{oot2012}.}}
\end{deluxetable*}

Previously, Centaur 29P had been observed with NEOWISE, Spitzer, and AKARI, all having the capability to observe CO$_2$ in comets.  However since NEOWISE and Spitzer detect CO and CO$_2$ in the same bandpass, there are no separate CO and CO$_2$ production rates provided from them.  AKARI had the capability to measure the CO$_2$ production rate, but only obtained an upper limit for 29P.  This makes our study of 39P the first CO$_2$ detection reported for a Centaur.  Upper limits for H$_2$O and CO show abundance ratios of Q$_{CO_2}$/Q$_{H_2O}$ $\geq$ 0.60 and Q$_{CO}$/Q$_{CO_2}$ $\leq$ 2.03 (Table \ref{tab:versus}).

\begin{figure}
\begin{center}
    \includegraphics[width=0.85\textwidth]{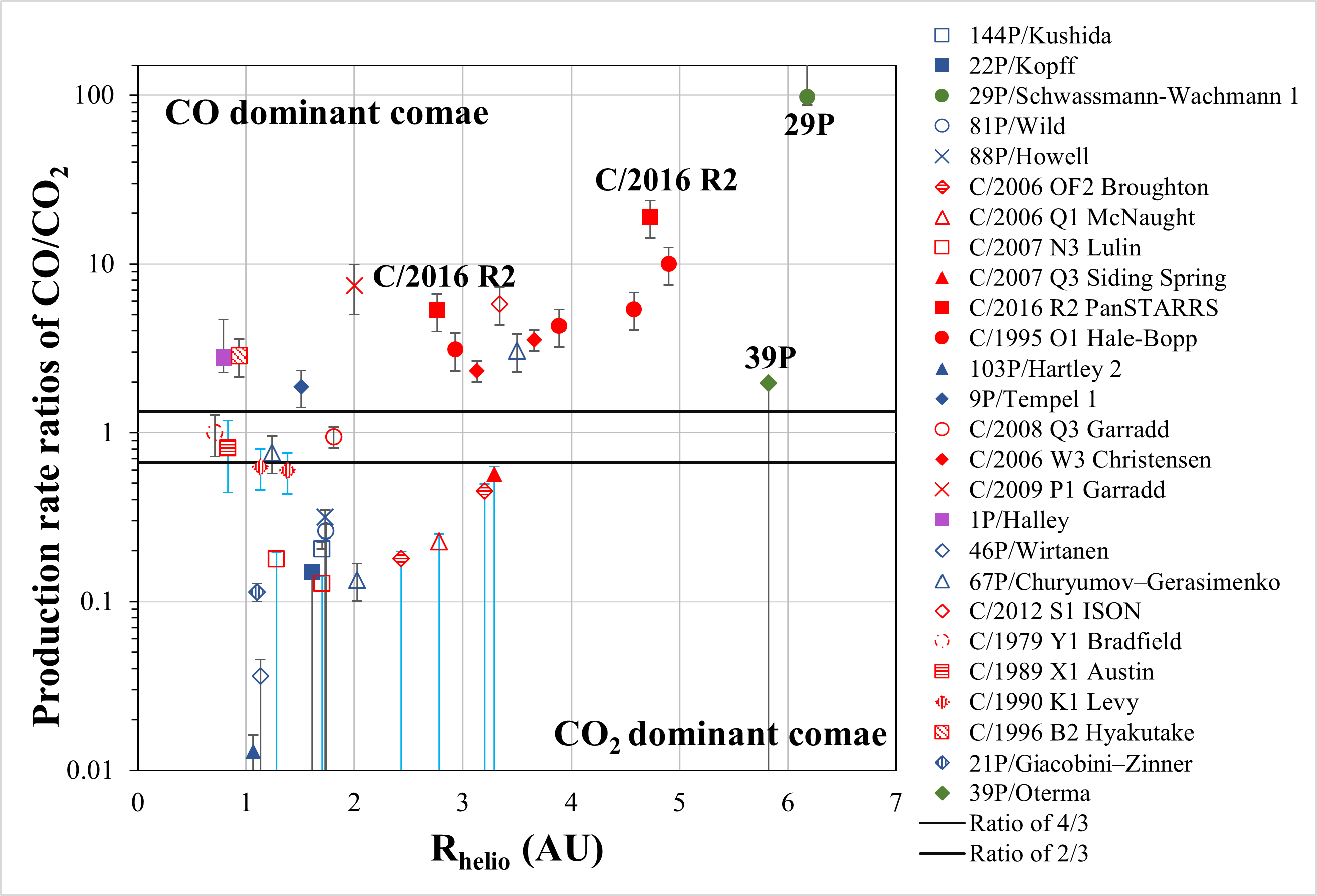}\hfill
    \caption{CO/CO$_2$ over different heliocentric distances.  39P is included as a green diamond for comparison to comets, and Centaur 29P (green circle).  The CO/CO$_2$ of 39P displayed in the graph is an upper limit.}\label{fig:39PCOCO2}
\end{center}
\end{figure}

\begin{figure}
\begin{center}
    \includegraphics[width=0.95\textwidth]{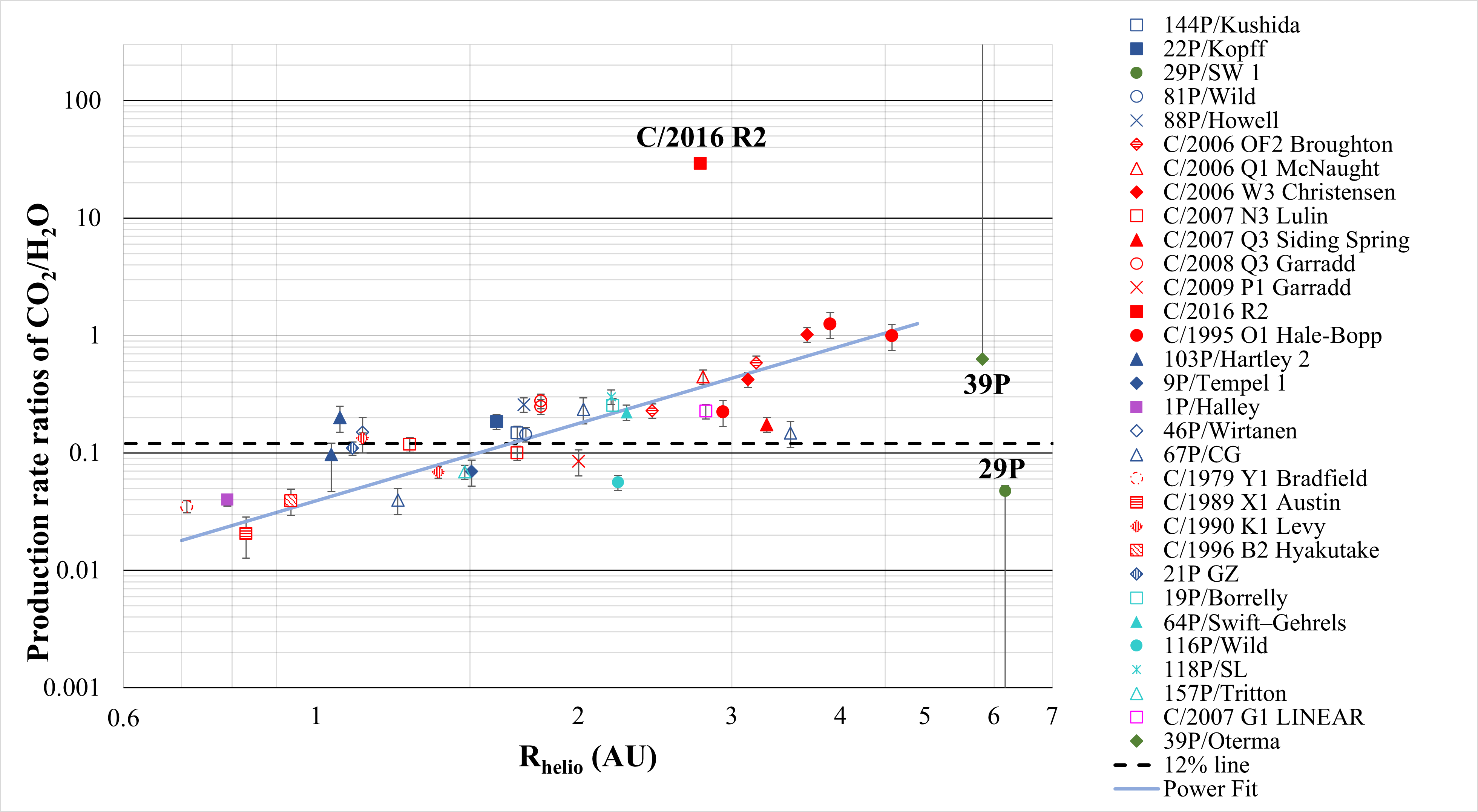}\hfill
    \caption{CO$_2$/H$_2$O over different heliocentric distances.  39P could possibly follow the fractional production rate of CO$_2$ to water.  39P is included, as a green diamond, to compare against comets, and Centaur 29P (green circle).  The CO$_2$/H$_2$O of 39P displayed in the graph is a lower limit. }\label{fig:39PCO2H2O}
\end{center}
\end{figure}

Although only upper limits were obtained for CO$_2$ in 29P from AKARI, the reported production rates of CO and H$_2$O for 29P show that it produces far more CO and H$_2$O than 39P (Figures \ref{fig:39PCOCO2} and \ref{fig:39PCO2H2O}).  The relative abundance ratios show that CO$_2$ production in 39P is greater than or equal to that of its CO, and that the CO$_2$ observed could be at least $\sim$ 40\% of the H$_2$O, while for 29P the CO$_2$ production is far less than either of these molecular species. 

With the difficulty of observing CO$_2$ from Earth-based telescopes, there are only a few telescopes intended as observatories that had the spectroscopic capability of measuring IR spectra of CO$_2$ -- ISO and AKARI \citep{HP2022} -- and prior to JWST, only one of those (AKARI) had attempted to observe it in a Centaur (29P).  Since so little Centaur data exist, we compare the relative abundance ratios of 39P against comets in Figs. \ref{fig:39PCOCO2} and \ref{fig:39PCO2H2O}.  In Figure \ref{fig:39PCOCO2} we see how 29P has a much larger amount of CO compared to CO$_2$, while 39P falls into the equal parts and CO and CO$_2$ region as well as the CO$_2$ dominant region, which is completely different from 29P and comets (Oort Cloud Comets) observed at that heliocentric distance ($\sim$ 4-6 au).  One explanation for the different behavior observed for 39P is the fact that its nucleus has undergone more dramatic temperature changes from drastic orbital changes in a shorter period of time than for 29P.  In Figure \ref{fig:39PCO2H2O}, 39P once again shows the opposite behavior of 29P, with more CO$_2$ versus water observed in 39P than the other Centaur or the median among comets.  This difference between these two Centaurs will be useful for constraining models of their formation and subsequent evolution.

The production rates derived from the 39P spectrum show a very different side to Centaurs so far, starting with the fact that only CO$_2$ was detected, and upper limits were obtained for its CO and H$_2$O production, while for 29P CO and H$_2$O were detected and an upper limit of CO$_2$ production was reported.  Figures \ref{fig:39PCOCO2} and \ref{fig:39PCO2H2O} are the modified graphs of Figure 2 and Figure 6 from \citet{HP2022}.  In Figure \ref{fig:39PCOCO2}, 39P is the only object in the 5-6 au region of the graphs, but the relative CO/CO$_2$ abundance ratio upper limit for 39P is very different from all the other objects.  39P is more CO$_2$ preferential than the CO dominant comets seen just below the 5 au line and 29P observed beyond the 6 au line.  The amount of CO$_2$ produced by 39P is at least comparable to twice the CO production or lower, which is much lower than distantly active comets and 29P and more consistent with comets observed at $r$ below 3.5 au. 

The fact that the relative production rates (or their significant limits) of CO/CO$_2$ and CO$_2$/H$_2$O have the opposite dominant molecular volatiles to those for 29P compared to 39P may suggest that Centaurs vary in composition, or that the behaviors of Centaurs change dramatically with heliocentric distance or processing history.  There are different contributing factors as to why there is a scarcity of data beyond 3.5 au.  One reason for a lack of comet/centaur observations in the CO$_2$ dominant regime at distances greater than 3.5 au is because this class of comets/Centaurs, with relatively inactive/very low CO$_2$ gas, have no detectable gas emissions without the sensitivity of JWST.  Another reason is possibly due to observational bias that large objects have more gas activity and make for a better candidate for observation, so small objects (like 39P) with low CO$_2$ would be tricky to measure without the capabilities of JWST. The increased solar heating that 39P experienced with its 3.4 au perihelion versus 5.5 au perihelion may have expended primordial hypervolatiles like CO and CO$_2$ \citep{Groussin2019,Gkotsinas2022}.  Since 39P has undergone significant changes in its orbit that 29P has not, this could result in the different abundance ratios.  Another reason that we see differences in the production rates could be indicative of a different initial chemical composition/formation region from 29P, or the different size of these centaurs could contribute to the different production rates. More data are needed to explore the possibilities, but this initial look at Centaurs is indicative that they are not all the same, and that future proposals and missions should look at the CO, H$_2$O, and CO$_2$ abundances of Centaurs.

\begin{figure}
\begin{center}
    \includegraphics[width=0.75\textwidth]{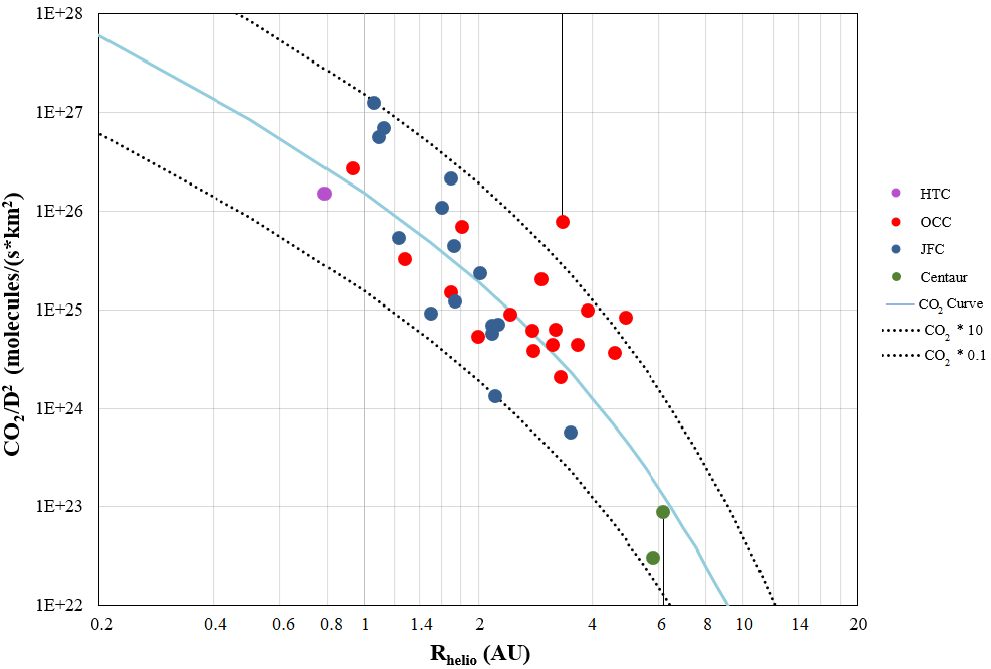}\hfill
\end{center}    
\caption{The production rates per unit area of 39P compared against other Centaurs and comets are shown.  The CO$_2$/D$^2$} is an updated version of Figure 12 from \citet{HP2022}.  39P shows low production rate per unit area, like the upper limit for 29P from \citet{oot2012}.\label{fig:ProdArea} 
    
\end{figure}

The active fraction of the volatiles to the size of an object can give an idea of level of activity an object is undergoing relative to its size.  In Figure \ref{fig:ProdArea}, comets and Centaurs are plotted against the CO$_2$ sublimation curve, to help explore activity correlations between these objects and relative to CO$_2$.  The production rates to unit area, seen in Figure \ref{fig:ProdArea}, show the low production rate to unit area of activity that was observed in 39P.  The CO$_2$/D$^2$, where D is the diameter of the nucleus and is based on a nominal $r$-band cometary geometric albedo of 0.05, was 2.6$\times$ 10$^{22}$ molecules s$^{-1}$ km$^{-2}$, and a CO/D$^2$ $\leq$ 2.5$\times$ 10$^{22}$ molecules s$^{-1}$ km$^{-2}$, using the radius from the Gemini data (see Section~\ref{sec:nuc-size}).  The CO$_2$/D$^2$ of 39P is in the range of the upper limit for the CO$_2$/D$^2$ of 29P, and the upper limit of CO/D$^2$ for 39P is among the lower production rates for its size.  This shows that these observations of 39P were taken in a time of low activity, relative to other objects, and the impressive capabilities of JWST were still able to capture the activity of 39P.\\

\subsection{Spatial profiles of CO$_2$ and dust}
\label{subsection:spdust}

\begin{figure}[h]
\begin{center}
\includegraphics [width=1\linewidth] {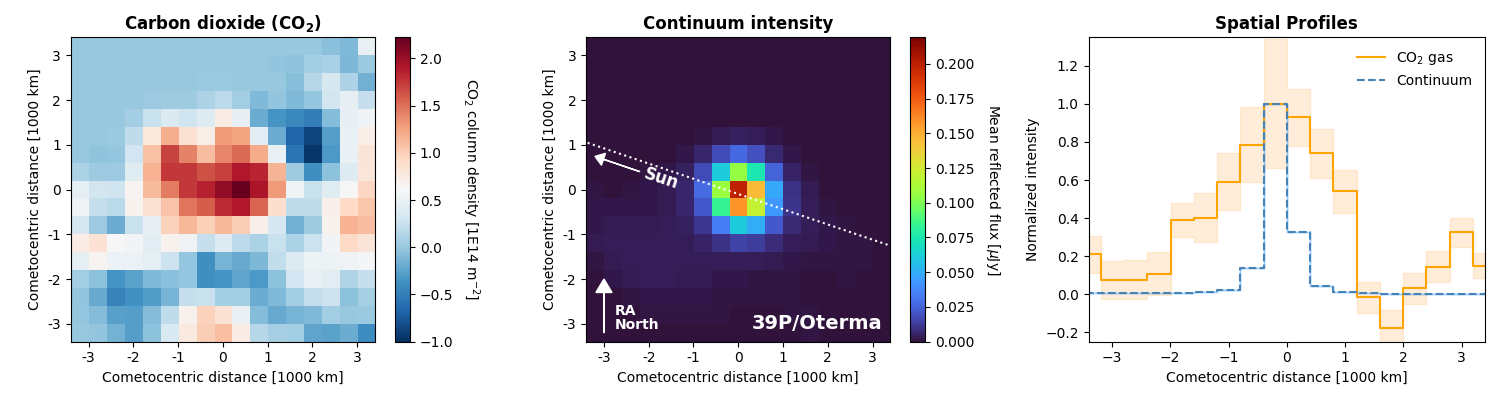}
\caption{ \label{fig:39P_CO2Profile} The CO$_2$ column density (left), continuum intensity (center), and their 1-D relative spatial profiles (right).  The spatial profile of CO$_2$ gas is broader than the dust profile, and it shows a possible small asymmetry in the sunward direction (dashed line in the second panel) when compared to the continuum (nucleus and dust) profile, which instead appeared centered and symmetric.  Both profiles are normalized to unity for comparison and visualization purposes.  The 1$\sigma$ uncertainties in the continuum profile are not as visible in the plot because of the higher SNR relative to the gas profile. 
}
\end{center}
\end{figure}

Spatial profiles are important since they can provide details on the behavior of the molecular and dust outflows relative to the center of the comet, where we assume the nucleus is located.  In Figure \ref{fig:39P_CO2Profile} we report 39P gas and dust maps, as extracted from the NIRSpec PRISM IFU.  To derive the CO$_2$ column density [m$^{-2}$] map, left panel, we first extracted the CO$_2$ flux map [$\mu$Jy] over the integrated band of CO$_2$ at 4.26 $\mu$m, and subtracted the continuum, to establish that what we are seeing is solely from the gas outflow.  The continuum underneath the CO$_2$ band, 4.19-4.29$\mu$m, was estimated as the mean value over 20 pixels extracted before and after the CO$_2$ band wavelength, 4.09-4.19$\mu$m and 4.29-4.39$\mu$m.  The CO$_2$ column density [m$^{-2}$] map was then obtained by adopting the fluorescence models from the Planetary Spectrum Generator \citep{Villanueva2018}.  In the middle panel of Figure \ref{fig:39P_CO2Profile} we report the mean reflected continuum intensity [$\mu$Jy], obtained by extracting the continuum at shorter wavelengths (between 0.7 - 2 $\mu$m); this represents the reflected sunlight from the nucleus and dust from 39P.
However being that the nucleus of 39P is a few km size (see Section \ref{sec:nuc-size}), and since 1 pixel corresponds to $\sim$ 398.94 km (0.1$\arcsec$), we can certainly consider the nucleus to be within the centered pixel of the maps, which points out to a more extended CO$_2$ gas outflow when compared to the mean reflected flux mapped from the continuum, which is more representative of the dust.

In the right panel of Figure \ref{fig:39P_CO2Profile} we show this comparison through their 1-D radial spatial profiles and relative 1$\sigma$ uncertainties, extracted over 25 pixels radius and centered on the nucleus.  We assume that the central peak pixel is the same for the gas and dust, and compare the profiles accordingly. The CO$_2$ gas profile seems to be slightly asymmetric and sunward-enhanced, when compared to the continuum profile.  This might be an indication of an extended coma related to an enhanced CO$_2$ outgassing.\\


\subsection{Ice absorption features}
\label{subsec:ice}

\begin{figure}
\begin{center}
    \includegraphics[width=0.495\textwidth]{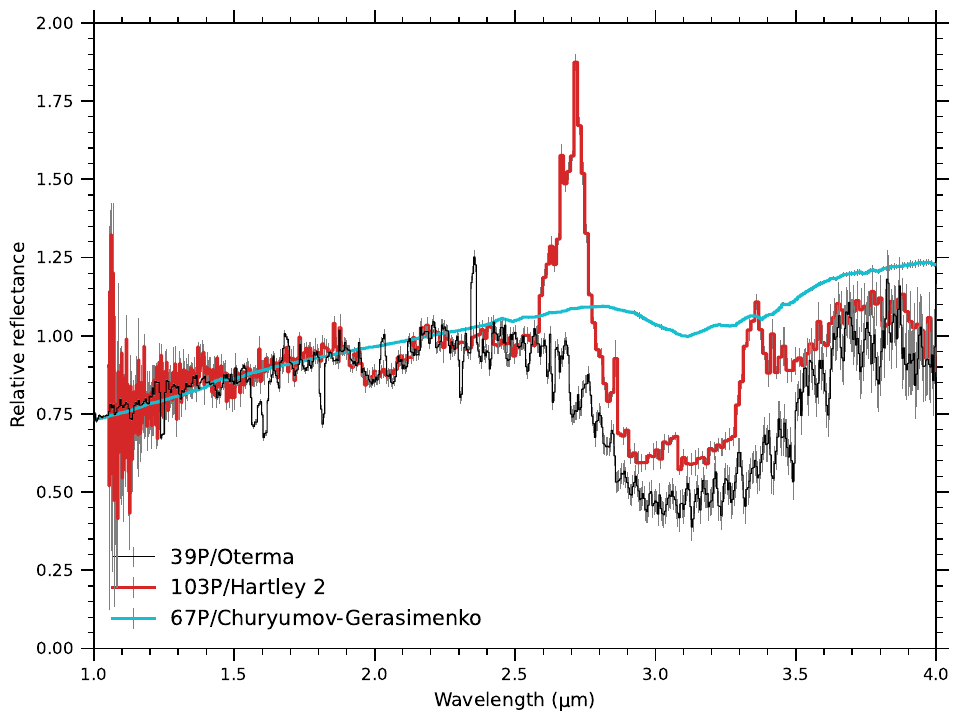}\hfill
    \includegraphics[width=0.495\textwidth]{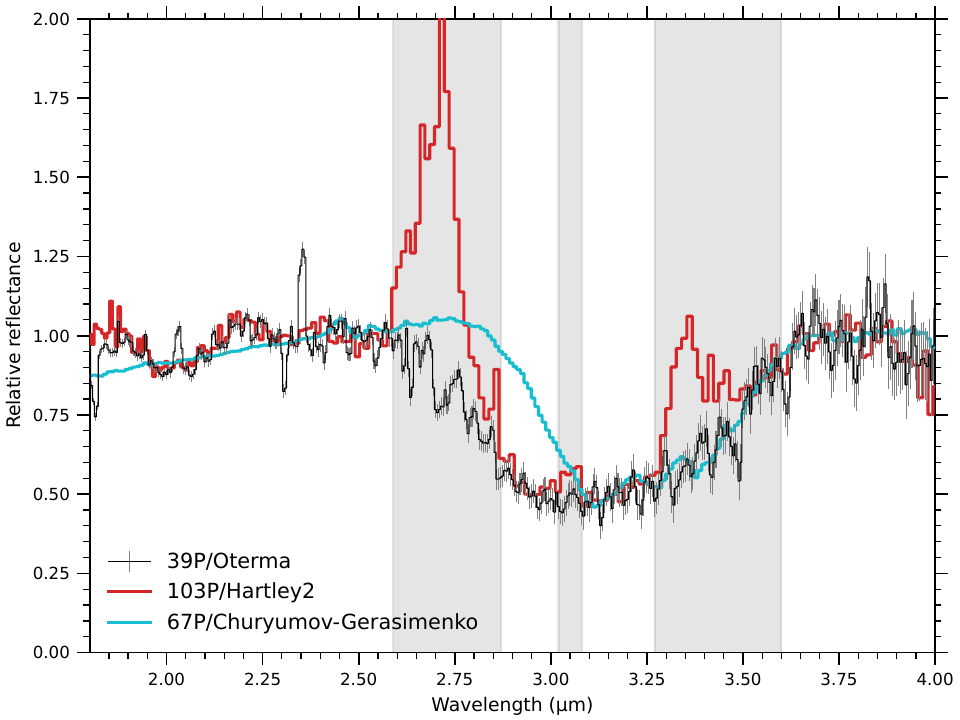}
    \caption{The continuum of Centaur 39P compared to the icy coma of comet 103P/Hartley 2 and the global average surface of comet 67P/Churyumov-Gerasimenko. (left) This graph investigates the 3-\micron{} bands, detrended and stretched as described in the text, we see that beyond 3.1~\micron 39P and 67P have the same spectral behavior.
    (right) This graph shows the relative reflectance spectra of all three objects (39P, 67P, and 103P), normalized to 1.0 at 2.2~\micron.\label{fig:continuum}}
\end{center}
\end{figure}

The relative reflectance spectrum of 39P is shown in Figure \ref{fig:continuum}.  The continuum exhibits absorption features at 2.0 and 3.1~\micron{} which we attribute to water ice.  For comparison, a spectrum of the icy coma of comet 103P/Hartley 2 \citep{protopapa14} and a spectrum of the global average surface of 67P/Churyumov-Gerasimenko are also shown \citep{raponi20-67p} in this figure. The three spectra in Figure \ref{fig:continuum} are showing different components of their respective comets (nucleus of 67P, coma of 103P, and an uncertain combination for 39P), which complicates interpretation. We are tacitly assuming that the composition and size distribution of icy grains in the coma match those properties on the nucleus, though that is not necessarily true since it is likely dependent on the particulars of surface structure, of recondensation, and of grain sublimation time scales for each comet. Furthermore, we know from many comets (and especially 67P) there is a diversity of surface behaviors.  With these caveats in mind, we proceed as follows.  All three spectra are normalized to 1.0 at 2.2~\micron{}, on the left of Figure \ref{fig:continuum}.  The continuum of all three objects are in excellent agreement from 1.0 to 1.9~\micron{}.  At 2.0-\micron{} both comet 103P and 39P show a shallow absorption feature, and at 3-\micron{} all three objects show an absorption feature.  However, the feature of 67P is much weaker than those of 103P and 39P, with band depths of 0.12, 0.45, and 0.50, respectively.

To further investigate the 3-\micron{} bands, we remove a linear spectral slope, based on the continuum at 2.25--2.55~\micron{} and 3.65--3.85~\micron.  The depths of the 103P and 67P bands are then scaled by 1.11 and 4.14, respectively, to match the depth of the 39P band at 3.1--3.2~\micron{} (this wavelength choice avoids a likely gas emission feature in the spectrum of 103P at 3.05~\micron).  The results are shown on the right of Figure~\ref{fig:continuum}.

The 67P spectrum is in good agreement with the 39P spectrum, but only longward of 3.1~\micron{}.  The short wavelength edges of the two comet bands do not match: 67P's band starts near 2.8~\micron{}, whereas 39P's band starts near 2.6~\micron{}. Furthermore, the 67P spectrum lacks 39P's 2-\micron{} band.  The shape of 67P's 3~\micron{} band has been attributed to aliphatic carbon and ammonium salts \citep{Poch2020,raponi20-67p}, but for 39P something else is needed to account for the spectral features.

In contrast with the 67P spectrum, there is excellent agreement between the shapes and relative depths of the 39P and 103P spectral features (except where gas emission bands are present).  The 2- and 3-\micron{} features of 103P are due to $\sim1$~\micron{} water ice grains \citep{protopapa14}.  The 2- to 3-\micron{} feature could be from small amounts of water ice in 67P \citep{raponi20-67p}.  We conclude that 39P has water ice grains with similar properties.  Cometary surfaces tend to have large grains $>10~\micron{}$, but comae have always shown small grains, $\lesssim1$~\micron{} \citep{protopapa18}.  This brings up the question of whether we are really seeing a surface or an unresolved coma.  Or, it may simply indicate the surface is covered with unusually small ice grains.  To resolve this dilemma may require applying the model of \citet{protopapa18} which can be explored in future work.

\section{Conclusions} \label{sec:summary}

Preceding this paper, CO$_2$ had rarely been detected in comets, and never in Centaurs.  The JWST observations of 39P on 2022 July 27 UTC, $r$ = 5.82 au, provide a CO$_2$ detection of (5.96 $\pm$ 0.80) $\times$ 10$^{23}$ molecules s$^{-1}$.  This is the lowest detection of CO$_2$ yet for any comet or Centaur, and is only possible due to capabilities of JWST NIRSPEC. While CO and H$_2$O were not detected, we obtained 3-sigma upper limits of Q$_{CO}$ $<$ 12.1 $\times$
10$^{23}$ molecules s$^{-1}$ and Q$_{H_2O}$ $<$ 10.0 $\times$ 10$^{23}$ molecules s$^{-1}$.

Using these significant upper limits for the abundance ratios leads to Q$_{CO}$/Q$_{CO_2}$ $\leq$ 2.03 for 39P while for 29P Q$_{CO}$/Q$_{CO_2}$ $\geq$ 83.2, which suggests that CO$_2$ is more dominant compared to CO in 39P versus 29P.  The Q$_{CO_2}$/Q$_{H_2O}$ $\geq$ 0.60 for 39P, while Q$_{CO_2}$/Q$_{H_2O}$ $\leq$ 0.05 for 29P, meaning that there is more CO$_2$ outgassing compared to H$_2$O in 39P versus 29P.  These relative production rate ratios show that 29P is not representative of all Centaurs. The opposite abundance ratios of 39P and 29P show that Centaurs could have very distinct behaviors. 29P should not be treated as "typical" Centaur until further data is collected to compare it against other Centaurs. If we compare these values to comets observed around that heliocentric distance, we see that 39P has no similarities in its CO/CO$_2$.  In terms of its CO$_2$/H$_2$O, it looks like 39P might follow the behavior of comets with an increasing CO$_2$/H$_2$O further from the Sun.  In any case, more observations are need to confirm any trend.

In the continuum of 39P, there are absorption features at 2.0 and 3.1~\micron{} which we attribute to water ice.
From the ice absorption features, we can see that the band depths for 39P closely match 103P, except where gas emission bands are present.  This suggests that the water ice grain size of 39P is similar to what \cite{protopapa14} measured in the coma of 103P: about 1~\micron{}.  This small grain size supports that the water ice being released from 39P and observed by JWST is in the coma of 39P rather than a detection of its surface \citep{protopapa18}, but a definitive conclusion cannot be made, given that the spatial profile of the continuum lacks a significant extended source.  As an alternative, the surface of 39P may be covered in widespread micrometer-sized ice grains. We also see that the continuum of 39P, 67P, and 103P are in excellent agreement from 1.0 to 1.9~\micron{} and at 3-\micron{} all three objects show an absorption feature.

Gemini and LDT observations of 39P have constrained the radius to $R_{nuc}= $2.21 to 2.49~km.  This is relatively small compared to the 32 km radius of 29P \citep{Schambeau2021} but similar to Centaurs like 423P/Lemmon, P/2005 S2 (Skiff), P/2010 TO20 (LINEAR-Grauer) and 2000 GM137 that have radii similar to the sizes of typical JFCs \citep{Jewitt2009, schambeau-2021CBET, schambeau-2023RNAAS}.  Both Gemini and LDT photometry and image analysis indicate there was no extended emission detected due to a conspicuous dust coma.  39P's detections were consistent with that of a point source at the time of observations, thus suggesting that we detected its bare nucleus in the $r^\prime$ filter, though at its geocentric distance we cannot rule out the presence of a diffuse coma.

\section{Acknowledgements} \label{sec:thanks}

We thank Quanzhi Ye for assistance with the Lowell Discovery Telescope observations.  OHP acknowledges the LSSTC Data Science Fellowship Program, which is funded by LSSTC from NSF Cybertraining Grant \#1829740, the Brinson Foundation, and the Moore Foundation; her participation in the program has benefited from this work.  This work is based in part on observations made with the NASA/ESA/CSA James Webb Space Telescope.  The data were obtained from the Mikulski Archive for Space Telescopes at the Space Telescope Science Institute, which is operated by the Association of Universities for Research in Astronomy, Inc., under NASA contract NAS 5-03127 for JWST.  These observations are associated with program \#2416.  Support for program \#2416 was provided by NASA through a grant from the Space Telescope Science Institute, which is operated by the Association of Universities for Research in Astronomy, Inc., under NASA contract NAS 5-03127.  This material is based on work supported by the National Science Foundation under grants AST-1615917 and AST-1945950.  This material is based in part on work done by MW while serving at the National Science Foundation. CAS acknowledges funding support from the Florida Space Research Initiative program.

These results made use of the Lowell Discovery Telescope (LDT) at Lowell Observatory.  Lowell is a private, non-profit institution dedicated to astrophysical research and public appreciation of astronomy and operates the LDT in partnership with Boston University, the University of Maryland, the University of Toledo, Northern Arizona University and Yale University.

The Pan-STARRS1 Surveys (PS1) and the PS1 public science archive have been made possible through contributions by the Institute for Astronomy, the University of Hawaii, the Pan-STARRS Project Office, the Max-Planck Society and its participating institutes, the Max Planck Institute for Astronomy, Heidelberg and the Max Planck Institute for Extraterrestrial Physics, Garching, The Johns Hopkins University, Durham University, the University of Edinburgh, the Queen's University Belfast, the Harvard-Smithsonian Center for Astrophysics, the Las Cumbres Observatory Global Telescope Network Incorporated, the National Central University of Taiwan, the Space Telescope Science Institute, the National Aeronautics and Space Administration under Grant No. NNX08AR22G issued through the Planetary Science Division of the NASA Science Mission Directorate, the National Science Foundation Grant No. AST–1238877, the University of Maryland, Eotvos Lorand University (ELTE), the Los Alamos National Laboratory, and the Gordon and Betty Moore Foundation.

This work was enabled by observations made from the Gemini North telescope, located within the Maunakea Science Reserve and adjacent to the summit of Maunakea.  We are grateful for the privilege of observing the Universe from a place that is unique in both its astronomical quality and its cultural significance.

Gemini results are based on observations obtained at the International Gemini Observatory (Prog. ID: GS-2023A-LP-203, PI: C. Schambeau), a program of NSF’s NOIRLab, which is managed by the Association of Universities for Research in Astronomy (AURA) under a cooperative agreement with the National Science Foundation on behalf of the Gemini Observatory partnership: the National Science Foundation (United States), National Research Council (Canada), Agencia Nacional de Investigación y Desarrollo (Chile), Ministerio de Ciencia, Tecnología e Innovación (Argentina), Ministério da Ciência, Tecnologia, Inovações e Comunicações (Brazil), and Korea Astronomy and Space Science Institute (Republic of Korea).

\facilities{JWST (NIRSPEC), Gemini North (GMOS), LDT (LMI CCD camera)}

\appendix
\section{Aperture correction for LDT data}\label{app:aperture}
The LDT images have a background source that passes 2\farcs2 away from our target comet.  We chose a small photometric aperture of 1\farcs2 to avoid contamination from the source, but due to the 1\farcs1 seeing, the photometry requires an aperture correction.  To derive the aperture correction for the LDT data, we rely on the radial profile of the Centaur itself, assuming it is azimuthally symmetric.  This assumption is justified by the JWST observations, which lack a dust coma.  Moreover, it avoids deriving an aperture correction from the trailed stars (trailing is 1\farcs2 per exposure).

\begin{figure}[H]
    \centering \includegraphics{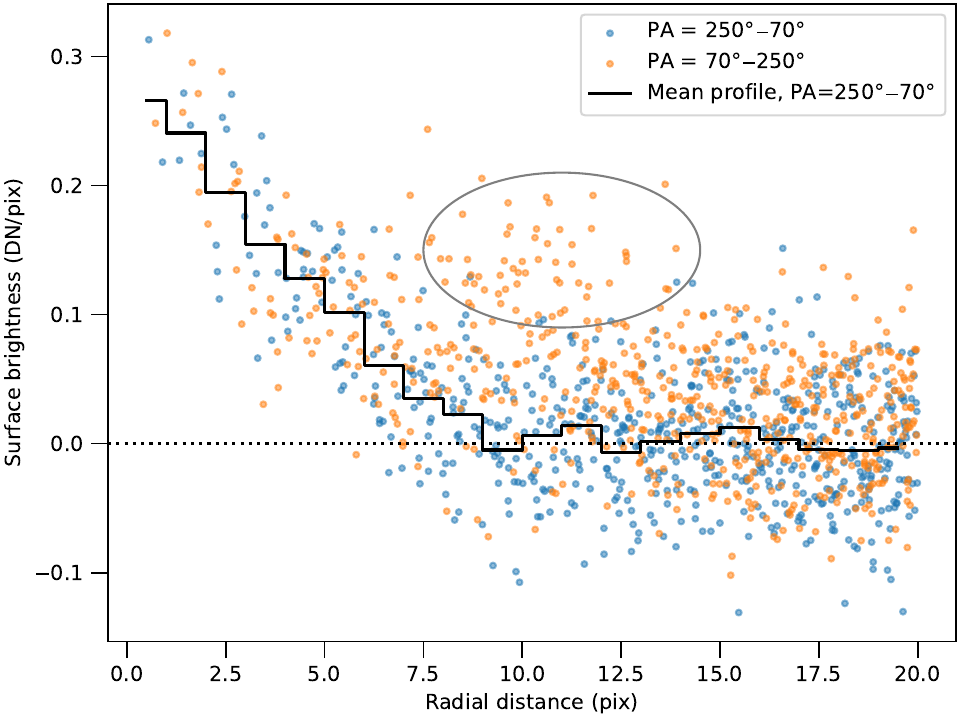}
    \caption{Radial profiles of 39P from LDT r$^\prime$ images.  The area around the Centaur is divided into two halves based on the position angle of the comet's motion (1\farcs2 = 5~pix per exposure along PA=250\degr), and the data points colored accordingly.  An ellipse highlights pixels within the PA=70\degr--250\degr{} half that are affected by the background source.  The mean radial profile for the unaffected half (PA=250\degr--70\degr) is shown as a line.}
    \label{fig:aperture-correction}
\end{figure}

Figure~\ref{fig:aperture-correction} displays the pixel-by-pixel radial profile of 39P based on an average of our three exposures.  The profile is divided by position angle, PA, into two halves: the area between 70\degr{} and 250\degr{} east of north is toward the contaminating source, and the area between 250\degr{} and 70\degr{} is clean sky away from the source.  The contaminating source itself is apparent as a surface brightness excess from $\sim$8 to 13~pix from the 39P centroid.  We also show the mean radial profile for PA=250\degr{} to 70\degr.

We measured the brightness of 39P in a 5-pix (1\farcs2) radius aperture, and the radial profile of the clean half reaches the background at 9-pix (2\farcs16).  The aperture correction is based on the integration of the mean profile, $\bar{I}(r)$, where $r$ is the distance from the 39P centroid:
\begin{equation}
    AC = \frac{\int_0^{11}\bar{I}(r) r dr}{\int_0^5\bar{I}(r) r dr}.
\end{equation}
The result, $AC = 1.670$, is a multiplicative factor for the measured photometry within 5~pix.

\bibliography{sample631}{}
\bibliographystyle{aasjournal}



\end{document}